\documentclass{edp-jp4}
\usepackage{graphicx}
\begin{document}

\title{Aging phenomena in polystyrene thin films} 
\author{K. Fukao}\address{Department of Polymer Science, Kyoto Institute of Technology, 
Matsugasaki,\\
Kyoto 606-8585, Japan}
\author{H. Koizumi}\sameaddress{1}
\maketitle
\small
\vspace*{-0.5cm}
\begin{abstract} 
The aging behavior is investigated for thin films of atactic polystyrene 
through measurements of complex electric capacitance.
During isothermal aging process 
the real part of the electric capacitance increases
 with aging time, while the imaginary part
decreases with aging time.
 This result suggests that 
 the aging time dependence of the real and imaginary parts are
 mainly associated with change in thickness and
 dielectric permittivity, respectively. In thin films, 
the thickness depends on 
thermal history of aging even above the glass transition. 
 Memory and `rejuvenation' effects are also
 observed in the thin films. 
\end{abstract}
%
\section{Introduction}
Polymeric glasses show structural changes during aging process 
below glass transition temperature $T_{\rm g}$, and corresponding 
changes in many physical quantities are 
observed~\cite{Struick,Bouchaud}. 
These phenomena known as physical aging are regarded as an 
important common property characteristic of
disordered materials including polymer
glasses~\cite{Bellon1,Fukao2,Fukao5} and spin
glasses~\cite{Lefloch,Vincent}. 
In our previous papers~\cite{Fukao2,Fukao5}, it 
is found for thin films of poly(methyl methacrylate) (PMMA) that the 
dielectric constant decreases with aging time during 
aging process, and 
that the thin films show memory and rejuvenation
that depend on thickness.

The glass transition dynamics of thin films of atactic polystyrene
(a-PS) have been investigated most intensively, and it has been reported that
the glass transition temperature $T_{\rm g}$ is drastically decreased
and the dynamics of the $\alpha$-process becomes faster with decreasing
film thickness~\cite{Keddie,Fukao1,Fukao3,Serghei,Lupascu}. 
Therefore, it should be an
interesting question whether the aging dynamics
below $T_{\rm g}$ can also be affected by the decrease in $T_{\rm g}$
in thin film geometry. 
In this paper, we investigate the aging phenomena for thin films of
a-PS through measurements of complex electric capacitance.

\section{Experiment}

Thin films of atactic polystyrene (a-PS) with the thickness $d$
of 14 nm and 293 nm (bulk) were prepared using a spin-coat method from a
toluene solution of a-PS on Al-deposited glass substrate. 
The samples of a-PS used in this study were purchased from 
Aldrich Co., Ltd. ($M_{\rm w}$=$1.8\times 10^6$, $M_{\rm w}/M_{\rm n}=1.03$). 
The preparation methods for thin films are the same as
in our previous papers~\cite{Fukao1,Fukao3}.
%
%
The values of $T_{\rm g}$ in thin films with $d$=14nm and 293nm are 
350 K and 370 K, respectively.

Capacitance measurements were done using an LCR meter (HP4284A) for the 
frequency $f$ from 20 Hz to 1MHz during the cooling and heating  
processes between 380K and 273K at a rate of 1K/min 
and also during isothermal aging at various aging temperatures
$T_a$ (=321.3 K $\sim$ 305.8K).
In our measurements, the complex electric capacitance of the sample
condenser $C^*(\equiv C'-iC'')$ was measured as a function of 
temperature $T$ and 
aging time $t$. The value of $C^*$ can be converted into 
the dynamic (complex) dielectric constant 
$\epsilon^*(\equiv \epsilon'-i\epsilon'')$ by 
dividing the $C^*(T)$ by the geometrical capacitance $C_0(T_0)$ 
at a standard temperature $T_0$.
The value of  $C^*$ is given by $C^*=\epsilon^*\epsilon_0\frac{S}{d}$ and
$C_0=\epsilon_0 \frac{S}{d}$, 
where $\epsilon_0$ is the permittivity in vacuum, $S$ is the area of 
the electrode. For evaluation of $\epsilon^*$ and $C_0$, 
we use the thickness $d$ which is determined at $T_0=$293 K and 
$S$=8 mm$^2$.


\section{Aging dynamics}

Fig.1 shows aging time dependence of $C'$ and $C''$ during the
isothermal aging for $d$=14nm at various values of $T_a$. 
In Fig.1(a), the deviation $\Delta C'$ of $C'$ from the value at the initial
time, at which the temperature of the sample reaches the aging
temperature, increases monotonically with aging time and the amount of
the relaxation for the isothermal aging at $T_a$ for 20 hours
decreases with decreasing $T_a$. In Fig.1(b), on the other hand, it is found
that $\Delta C''$ decreases with increasing aging time 
and the amount of the relaxation of $C''$ decreases with decreasing $T_a$. 
Comparing the results observed for $d$=293nm with those for $d$=14nm, 
it is found that the time dependence of $\Delta C'$ 
($\Delta C''$) for $d$=14nm is similar to that for $d$=293nm,
the relaxation strength depends on film thickness. 
As for the frequency dependence, we could observed that as $f$
increases, $\Delta C''$ decreases, while $\Delta C'$ remains almost 
constant. 
The detailed dependence on $d$ and $f$ will be reported elsewhere.


\begin{figure}
\includegraphics*[width=7.2cm]{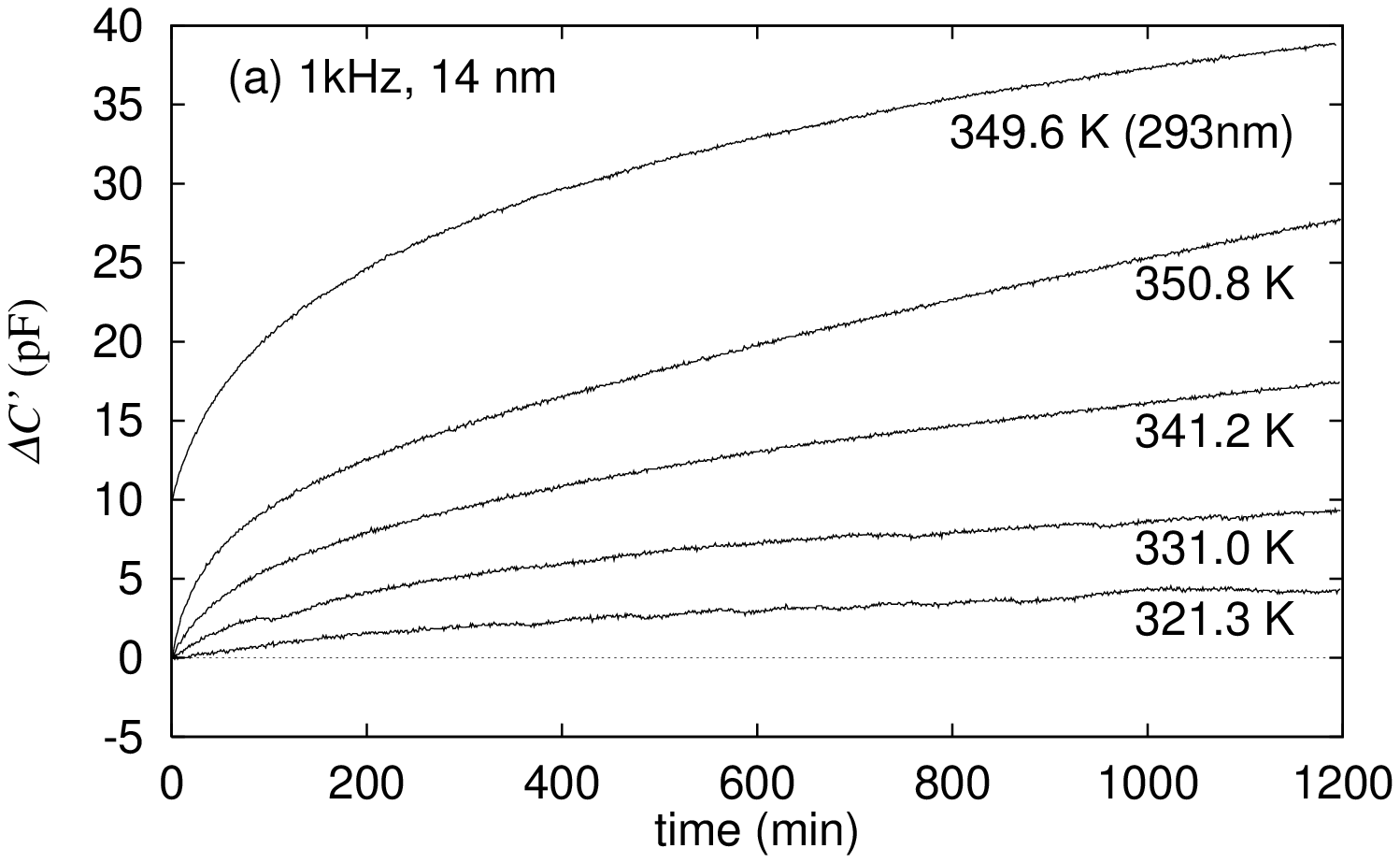}
\includegraphics*[width=7.2cm]{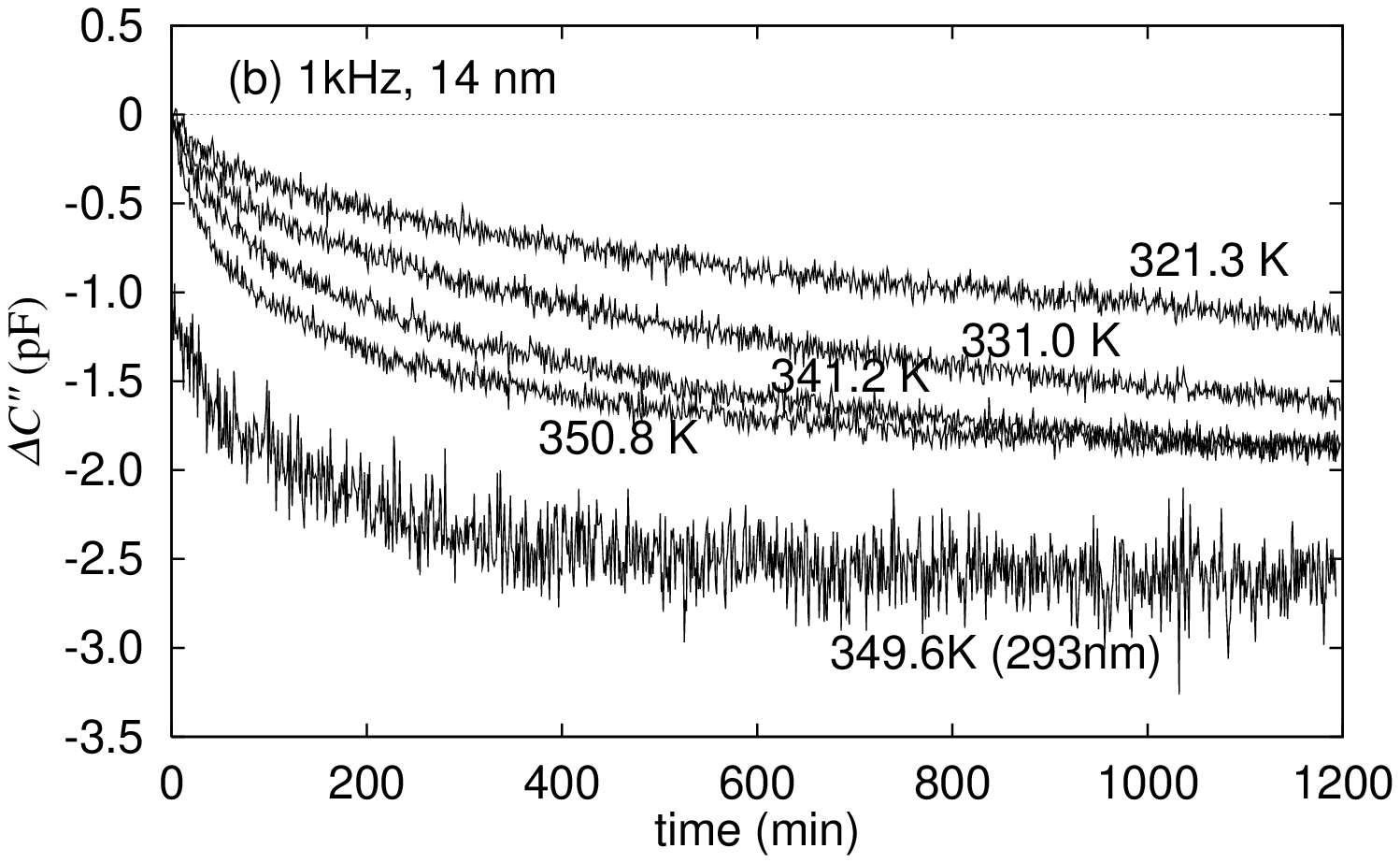}
\caption{\label{fig:2} Aging time dependence of the deviation 
$\Delta C'(t)$ ($\Delta C''(t)$) of $C'(t)$ 
($C''(t)$)  from the initial values $C'(0)$ ($C''(0)$)  for various 
aging temperatures for thin films of a-PS with film thickness of 14nm
 for $f$=1kHz. The aging temperatures $T_a$ are 350.8K, 341.2K, 331.0K and
 321.3K. As a reference, the results for $d$=293nm are also shown after
 rescaling and shifting along the vertical axis. 
} 
\end{figure}

Here, an explanation for the observed dependence of $C'$ and
$C''$ on the aging time at an isothermal aging process is given in the
following way. The real and imaginary parts of the complex electric
capacitance are given as:
\begin{eqnarray}
\label{Eq1}
C'(\omega,T,t)&=&(\epsilon_\infty (T,t)+\epsilon'_{\rm
 disp}(\omega,T,t))C_0(T,t)\\
\label{Eq2}
C''(\omega,T,t)&=&\epsilon''_{\rm disp}(\omega,T,t)C_0(T,t),
\end{eqnarray}
where $\epsilon_\infty$ is the dielectric constant at high frequency
limit, $\epsilon'_{\rm disp}$ and $\epsilon''_{\rm disp}$ are
frequency-dependent
contributions to dielectric constant due to orientational polarization
associated with molecular motions. Here, there are relations as follows:
$\epsilon'=\epsilon'_{\rm disp}+\epsilon_{\infty}$ and
$\epsilon''=\epsilon''_{\rm disp}$.  
In the case of a-PS, because the polarity is very week, it can be
expected that $\epsilon'_{\rm disp}\ll\epsilon_{\infty}$. Therefore, 
Eq.(\ref{Eq1}) can be rewritten approximately as 
\begin{eqnarray}
C'(\omega,T,t)&\approx&\epsilon_\infty(T,t) C_0(T,t)
\end{eqnarray}
For an isothermal aging process, it is expected that the density increases with
aging time, and hence $d$ decreases, on condition that $S$ 
remains constant. This density change, therefore, causes the
increase both in $\epsilon_{\infty}$ and $C_0$, as a result, in $C'$ 
according to the discussion given in 
Ref.\cite{Fukao1}. 

On the other hand, Eq.(\ref{Eq2}) shows that $C''$ includes two different
 contributions from $\epsilon''_{\rm disp}$ and $C_0$. For the isothermal
 aging, $C_0$ should increase with aging time, because the density increases,
 $i.e.$, the film thickness decreases, as shown in the above. 
If $\epsilon''_{\rm disp}$ decreases with aging time, the
 contribution from $\epsilon''_{\rm disp}$ can compete with that from
 $C_0$, and there will be a possibility that $C''$ decreases with 
increasing aging time. 
For PMMA, which has a strong polar group within a chain, it has been
 reported that both $\epsilon'$ and $\epsilon''$
 decrease with aging time during isothermal aging process.
If we assume that $\epsilon''_{\rm disp}$ decreases with increasing aging time 
for a-PS in the similar way
as observed in PMMA, and that the decrease in $\epsilon''_{\rm disp}$ 
overcomes the increase in $C_0$, $C''$ can be
decreased with increasing aging time. From the above discussions, the
results observed in a-PS can be explained as follows: for the
isothermal aging process, {\it the decrease in $d$} is observed as
the increase in $C'$, while {\it the decrease in $\epsilon''$} is 
observed as the decrease in $C''$. Therefore, the
present measurement 
will give us information
on the change in volume and dielectric permittivity simultaneously for 
the same sample during the isothermal aging process.

\section{Memory and `rejuvenation' in thin films}

\begin{figure}
\hspace*{0.5cm}\includegraphics*[width=6cm]{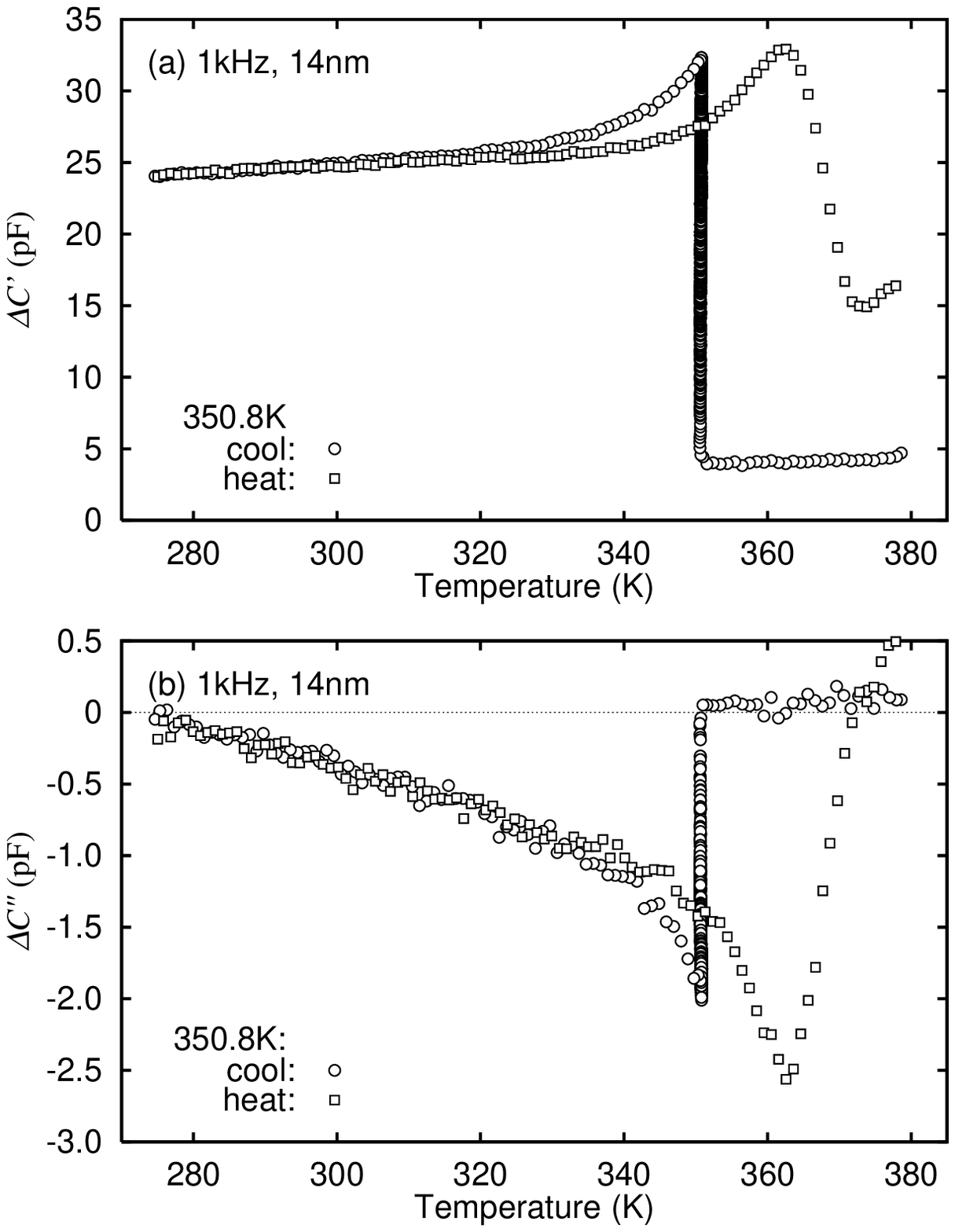}
\includegraphics*[width=6cm]{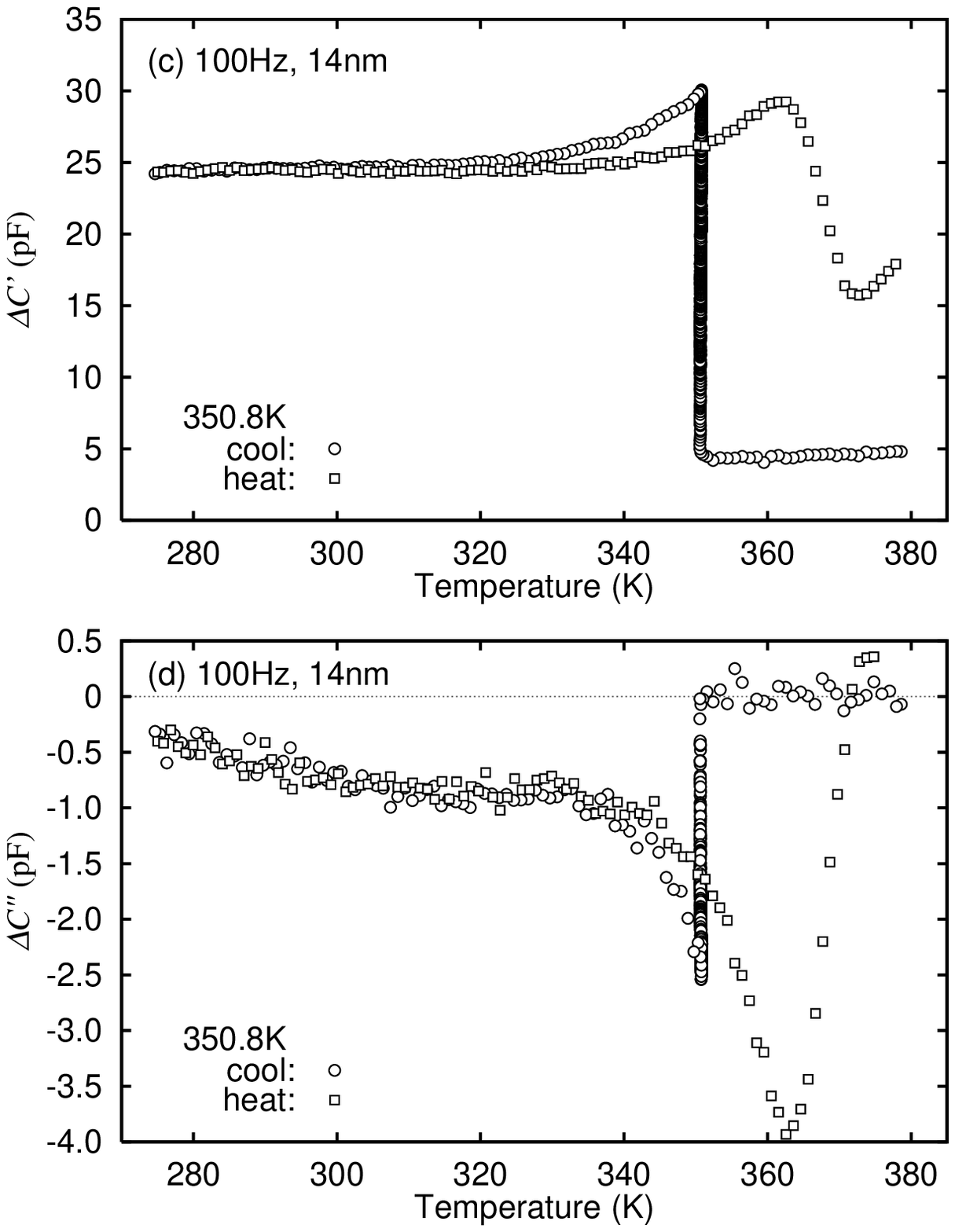}

\caption{\label{fig:3} 
Temperature dependence of the deviation $\Delta C'$ ($\Delta C''$) of 
the component of the complex electric capacitance $C'$ ($C''$) from the 
reference values observed for
the cooling process (circles) including at 
isothermal aging at 350.8K and the subsequent heating process (boxes)
for thin films of a-PS with $d$=14nm and $f$=1kHz ((a), (b))
and 100Hz ((c), (d)).}
\end{figure}

Fig.2 shows temperature dependence of $\Delta C'$ and $\Delta C''$
observed during the cooling process including isothermal aging at $T_a$
and the subsequent heating process for $f$=1kHz and 100Hz. 
In this case, $\Delta C'(T)$
($\Delta C''(T)$) are evaluated as the deviation of $C'(T)$ ($C''(T)$)
from reference values $C'_{\rm ref}(T)$ ($C''_{\rm ref}(T)$).
As the
reference value of the cooling (heating) process,
we used the data measured for the preceding cooling (heating) process
without any isothermal aging.

Fig.2(a) shows that as the temperature decreases from 380K to 350.8K,
$\Delta C'$ remains almost constant, and then 
$\Delta C'$ becomes larger as the aging time increases 
during the isothermal aging at 350.8 K. 
During the cooling process after the isothermal aging, the deviation
becomes smaller and then approaches to a constant value, but {\it not to 
zero}. As a result, most part of the
deviation $\Delta C'$ induced during the isothermal aging remains even
at 273 K. This result may be associated with the fact that the
isothermal aging increases the density.
For the subsequent heating process, $\Delta C'$ changes
along the path traced by $\Delta C'$ for the preceding cooling
process after the isothermal aging, and then $\Delta C'$
shows a maximum just above $T_a$. 
After that $\Delta C'$ rapidly decreases approaching a value above zero. 
This behavior can be interpreted as follows: the fact that the sample
experiences the aging at $T_a$ by the way of the cooling process is
memorized within the sample, and the memory is recalled during the
subsequent heating process. Here, it should be noted that there are two
interesting behavior. First, we find that there
is a small deviation from zero between 350.8K and 380K for the cooling
process.  Secondly, there is an
appreciable difference between $\Delta C'$ at 380K before and after the
temperature cycle, in which temperature changes from 
380K $\rightarrow$ 273K $\rightarrow$ 380K.  
If the sample is in an equilibrium state 
above $T_{\rm g}$, both the deviations should vanish. 
Furthermore, the existence of such deviations above $T_{\rm g}$ 
is not observed for the bulk sample ($d$=293 nm).
This result implies that thin films of a-PS are
not in the equilibrium state even above $T_{\rm g}$, which may be
related to the existence of a slow relaxation process of film thickness
in ultra thin films~\cite{Kanaya1,Reiter1}.

The temperature dependence of $\Delta C''$ is
different from that of $\Delta C'$, as shown in Fig.2(b).
For the cooling process from 380K, $\Delta C''$ remains almost zero,
and then $\Delta C''$ decreases with aging time for the
isothermal aging. After that, $\Delta C''$ increases with decreasing
temperature and reaches zero at about 273K, which suggests that the
system is `{\it rejuvenated}' as for the dielectric response.  

Combining the results observed for $\Delta C'$ and $\Delta C''$, it is
concluded that the volume of thin films of a-PS becomes smaller during
the isothermal aging, and this deviation from the reference value is
kept even below the aging temperature. On the other hand, dielectric
permittivity also becomes smaller during the isothermal aging, and the
deviation of the dielectric permittivity from the reference value 
is totally `rejuvenated' at lower temperature. 
The existence of the volume change observed in
the present measurement is consistent with the fact that there are
several reports relating to the change in volume or density due to 
physical aging. Although the system is not rejuvenated judging only 
from the volume of the system, the dielectric response to 
electric field is fully `rejuvenated'.

Fig.2(d) shows the temperature dependence of $\Delta C''$ for the
same temperature change for $f$=100Hz. In this figure, it is found
that $\Delta C''$ does not increase smoothly as the temperature changes 
from $T_a$ to
273K after the isothermal aging, but there is a plateau between
330 K  to 300 K for $f$=100Hz. 
This plateau region can be observed only in the ultra thin films, not 
in the bulk sample.
The position of the plateau almost conincides with that of 
an additional relaxation process, which can be observed only in thin
films and is referred to as the $\alpha_l$-process~\cite{Fukao1}. 
It is expected that the molecular motion
corresponding to the $\alpha_l$-process may be due to the segmental motion
in a mobile region in thin films. 
In Fig.2(d), there might be a possibility that the
`rejuvenation' is suppressed by the existence of the $\alpha_l$-process 
during the cooling process after the isothermal aging for $f$=100Hz.
It is expected that detailed investigations on the relation between 
the $\alpha_l$-process and the rejuvenation effect will lead to 
understanding the mechanism of the memory and `rejuvenation' effects 
in thin film geometry.

\section*{Acknowledgements}

The authors appreciate K.Takegawa and Y.Saruyama for useful
collaboration.
This work was supported by a Grant-in-Aid for Scientific Research
(B) (No. 16340122) from Japan Society for the Promotion of Science and
for Exploratory Research (No. 16654068) from the Ministry 
of Education, Culture, Sports, Science and Technology of Japan.


\end{document}